\documentstyle[epsfig]{aipproc}
\input epsf.tex   

\def \gta {\mathrel{\vcenter
     {\hbox{$>$}\nointerlineskip\hbox{$\sim$}}}}
\def \m{\ifmmode M_\odot\else M$_\odot$\fi}
\def \r{\ifmmode R_\odot\else R$_\odot$\fi}
\def \l{\ifmmode L_\odot\else L$_\odot$\fi}

\def\gmcm3{gm~cm$^{-3}$}
\def\gm-s{gm~s$^{-1}$}
\def\cm3s{cm$^3$~s$^{-1}$}

\def\erg-s{erg~s$^{-1}$}

\def\beq{\begin{equation}}
\def\eeq{\end{equation}}
\def\ref{\reference}
\def\gr{$\gamma$-ray}
\def\grs{$\gamma$-rays}
\def\grb{$\gamma$-ray burst}
\def\grbs{$\gamma$-ray bursts}
\begin{document}
\title{The Type Ib/c Supernova, Gamma-Ray Burst, Soft Gamma-ray
Repeater, Magnetar Connection}
\author{J. Craig Wheeler$^*$, Peter H\"oflich$^*$, Lifan Wang$^*$
and Insu Yi$^{\dagger}$}
\address{$^*$Department of Astronomy, The University of Texas
Austin, Texas 78751\\
$^{\dagger}$Korea Institute for Advanced Study, Seoul, Korea }
\maketitle
\begin{abstract}
The polarization of core-collapse supernovae shows that many if not
all of these explosions must be strongly bi-polar.  The most obvious
way to produce this axial symmetry is by the imposition of a jet
as an intrinsic part of the explosion
process.  These jets could arise by MHD processes in the formation
of pulsars and be especially strong in the case of magnetars.  The
jets will blow iron-peak material out along the axes and other
elements from the progenitor along the equator, a very different
composition structure than pictured in simple spherical ``onion skin"
models.  In extreme cases, these processes could lead to the
production of \grbs\ powered by strong Poynting flux.
\end{abstract}

\section{Polarized Supernovae and Jets}
We have found that all core-collapse events, 
supernovae of Type II and Type Ib/c, are polarized at the 1\% level
and some much more so (Wang et al. 1996; Wang et al. 1999).  
Our data suggest a very important trend: the smaller the hydrogen envelope
and the deeper within the ejecta we see,
the larger the observed polarization.  
Polarization of the level we observe then forces us to abandon timid phrases
like ``asymmetric supernovae."  For these events, 
it is appropriate to talk about ``bi-polar supernovae." 

The next issue is thus how to account for the observed high
levels of polarization.  An asymmetric impulse in an otherwise
spherical configuration will tend to turn spherical before homologous
expansion is reached by the propagation of lateral pressure 
gradients.  What is needed is the directed
flow of energy and momentum in a single direction for a time
that is substantial compared to the dynamical timescale.  
We need a jet.  This conclusion
is independent of any connection to \grbs, but, of
course, the potential for this connection is clear.

A preliminary study in which
conditions were selected to represent the sort of MHD jet
found by LeBlanc \& Wilson (1970; see also M\"uller \&
Hillebrandt 1979; Symbalisty 1984) has been presented by Khokhlov et al. (1999;
see also H\"oflich, Wheeler \& Wang 1999).
This study has been extended to explore a range of jet energies
and stellar configurations, both bare helium cores and
red supergiants.

The code employed, developed by Khokhlov (1998), is an Eulerian
adaptive mesh code.  The
calculations are fully three dimensional.  The adaptive mesh gives
great resolution.  The finest scale corresponds to a uniform grid
of some $10^{10}$ cells.  The adaptive mesh also allows substantial
dynamic range. For the jet models this ranges from $2^{12} \sim 10^4$ to
$2^{19} \sim 10^6$.  The imposed jets are cylindrically symmetric
and the initial stellar model is spherical.  The resulting jets are
thus highly cylindrically symmetric, but this is not imposed in
the dynnamics, only the initial conditions.  The jet dynamics are
sufficiently rapid for the models computed that, Kelvin/Helmholz
instabilities have little time to form.

Figure 1 shows the distribution of the jet matter (unspecified in the
computation, but presumably rich in iron-peak elements), and of
the oxygen layers of the star.  The former reflects the bi-polar nature
of the jet flow. The latter shows the effects of the lateral shocks that
compress the oxygen into an equatorial shell.  This will, in turn,
affect the line profiles of the oxygen observed in the
nebular phase.  These profiles are presented after 4.84 seconds when
the jet breaks through the surface of the helium core.  They must
be followed into homologous expansion before any direct connections
to observations can be made.
We have also studied models with red giant envelopes (H\"oflich,
et al. 2000).  The code
allows us to follow the jet in a single calculation from the
center of the star out through the extended envelope.  We find
that energetic jets can penetrate the hydrogen envelope, but
that more modest jets cannot.  The latter can still induce
an asymmetric, bi-polar explosion.


\begin{figure}[t]
\begin{minipage}[t]{2.5truein}
\mbox{}\\
\epsfig{file=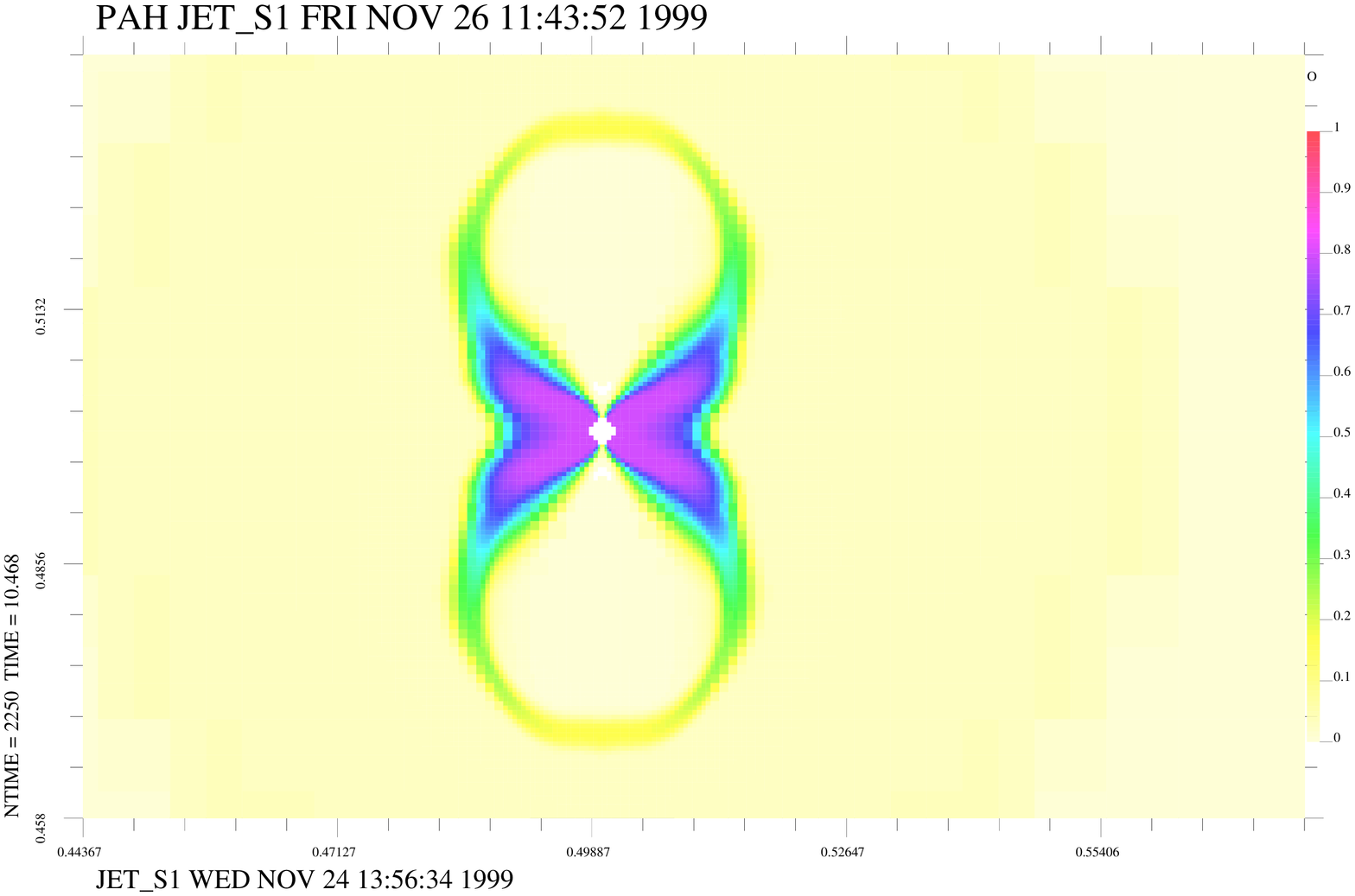,height=2.5truein,width=2.5truein,
clip=,angle=270}
\end{minipage}
\hfill
\begin{minipage}[t]{2.5truein}
\mbox{}\\
\caption{
Composition structure of a jet-driven supernova.  The axial jet
(light lobes) contains jet material.  The equatorial
shell (darker region) shows the distribution of the oxygen layer from
the initially spherical progenitor model (from H\"oflich et al. 2000).
}
\end{minipage}
\label{figure 1}
\end{figure}                 

The next issue is the origin of the jets.
To account for normal
supernovae we must have jets in routine circumstances, that
is, associated with the formation of a neutron star and not restricted to
the more rare circumstances of the possible formation of
a black hole.  This statement is independent of
the liklihood that in rare cases or different circumstances
such a jet might yield a \grb.

The obvious place to look for jets in frequent core
collapse events is in the rotating,
magnetic collapse of a neutron star with the equivalent dipole
magnetic field ranging from ``typical" values like the Crab
pulsar to the extreme values associated with magnetars
(Duncan \& Thompson 1992)
and soft \gr\ repeaters (Kouveliotou et al. 1998).
This environment gives a framework in which to
quantitatively address questions of physics that are
germane to the nature of the core collapse process in general and to
potential \gr\ production.
The physics that could be at play
in such a collapse has recently been considered by Wheeler
et al. (1999).

Rotation and magnetic fields have a strong potential to create
axial matter-dominated jets that will drive strongly asymmetric
explosions for which there is already ample observational
evidence in Type II and Type Ib/c supernovae, their remants,
and in the pulsar velocity distribution.
The potential to also create strong flows of Poynting flux
and large amplitude electromagnetic waves (LAEW)
serves to reinforce the possibility to generate bi-polar explosions.
These bi-polar explosions will, in turn, affect nucleosynthesis and issues such
as fall-back that determine the final outcome to leave behind
neutron stars or black holes. In addition, the presence of
matter-dominated and radiation-dominated jets might lead
to bursts of \grs\ of various strengths.  The issue of the
nature of the birth of a ``magnetar" in a supernova explosion
is of great interest independent of any connection to \grbs.
Highly magnetized neutron stars might represent one out of
ten pulsar births.  Production of a strong \grb\ might be
even more rare.

Wheeler et al. show that the contraction phase of a proto-neutron star
could result in a substantial change in the physical properties
of the environment.
When the rotating magnetized neutron star first forms there is
likely to be linear amplification of the magnetic field and
the creation of a matter-dominated jet, perhaps catalyzed by MHD effects,
up the rotation axis.  The rotational energy
of the proto-neutron star is $\sim 10^{51}$ ergs,
sufficient to power a significant
matter jet, but unlikely to generate a strong \grb.  The matter jet could
generate a smaller \grb\ as seems to be associated with
SN~1998bw and GRB~980425 by the Colgate (1974) mechanism as
it emerges and drives a shock down the stellar density gradient
in the absence of a hydrogen envelope, e.g., in a Type Ib/c supernova.

As the neutron star cools, contracts,
and speeds up, the rotational energy increases.  The energy becomes significantly
larger than required to produce a supernova and sufficient,
in principle, to drive a cosmic \grb\ if the collimation
is tight enough and losses are small enough.
For a neutron star with a period near 1 millisec the rotation
energy becomes $\gta 10^{52}$ ergs.  If efficiently utilized and
collimated, this energy reservoir could make a substantial
\grb.   The luminosity is estimated to be $\sim 10^{52}$ \erg-s\
and to last for a few seconds.

The second important factor the accompanies the contraction
and spin-up of the cooling neutron star is that the light cylinder contracts
significantly, so that a stationary dipole field cannot form
and the emission of strong LAEW occurs.  Tight collimation
of the original matter jet and of the subsequent flow of LAEW
in a radiation-dominated jet is expected.
The LAEW will propagate as intense low frequency, long
wavelength radiation.  
The LAEW ``bubble" could be strongly Rayleigh-Tayor unstable,
but still may propagate selectively with small opening angle
up the rotation axis as an LAEW jet.  
If a LAEW jet forms, it can drive shocks which may selectively propagate
down the axis of the initial matter jet or around the perimeter of
the matter jet.  The shocks associated with the LAEW jet could generate
\grs\ by the Colgate mechanism as they propagate down
the density gradient at the tip of the jet or there
could be bulk acceleration of protons to above
the pion production threshold.  The protons could produce
copious pions upon collision with the surrounding wind,
thus triggering a cascade of high energy \grs, pairs,
and lower-energy \grs\ in an observable \grb.
Yet another alternative is that the LAEW
could eventually propagate into such a low density
enviroment that they directly induce pair cascade.
The energy produced by the spin-down of the pulsar could emerge from
the stellar surface along the axis of a low-density matter jet, or in
an annulus surrounding a high density jet.  Either of
these cases will give a Lorentz factor that depends strongly
on the aspect angle of the observer.  

\section{Conclusions}

Circumstantial evidence has accumulated that the \grb\ phenomenon
is linked to Universal star formation and hence to massive
stars.  This alone does not say whether the product of the
massive star is a black hole or a neutron star.  Other evidence
suggests that there may be a roughly canonical energy in
\grs\ $\sim10^{52}$ ergs that may appear as a larger
``isotropic equivalent" energy in some cases because of collmation 
effects.  If this remains a relevant number, then the possible
association of some \grbs\ with neutron stars is still
on the table.  This energy is about what would expect in
the rotation of a newly formed neutron star, and it could
be delivered up in the form of Poynting flux in a few
seconds if the neutron star is very highly magnetized; if
it is a magnetar.
Two key facts emerge that might support the connection of
some supernovae that make neutron stars with some \grbs.
\newline\noindent
$\bullet$ Core collapse supernovae are strongly polarized.
\newline\noindent
$\bullet$ Magnetars exist!
\newline\noindent
This means that routine neutron star, not just black hole
formation requires the production of strong jets that
may themselves explode the star.  In addition,
we must understand the birth event of strongly magnetized neutron
stars.

Consideration of the systematics of of the formation of magnetars
suggests that the rotating collapse will first launch an
MHD jet up the rotation axis.  Later, after the neutron star
cools and contracts, it will generate a Poynting flux that
could be very intense for magnetar-type field strengths.  That
Poynting flux could emerge as a radiation-dominated jet following 
the path of the first matter-dominated jet.  Because the 
radiation-dominated jet cannot form for several seconds as
the neutron star contracts, spins up, and generates a large
magnetic field, but then propagates faster than the matter-
dominated jet, the matter jet could precede or follow
the radiation dominated jet.  In the former case an X-ray
precursor could be generated.  In the latter case the matter
jet might not be conspicuous at all.

The process of neutron star spin-down has predictable
properties if a simple dipole magnetic field is assumed.
The bolometric luminosity declines like $L_{bol} \propto
t^{-2}$ (Wheeler et al. 1999) and Blackman \& Yi (1998)
have estimated that for a synchrotron/Compton model
the luminosity in the BATSE band might scale like
$L_{BATSE} \propto t^{-1}$.  The ratio of the energy
emitted after several tens of seconds to the total
energy is a few percent.  These decline rates and the
efficiency are very reminiscent of the ``tails" of
\grbs\ described at this meeting by Litvine, Connaughton,
and Giblin.  After the meeting, Giblin (private communication)
reported that there is no sign of a periodic signal in
the data from the especially bright tail source GRB~980923.
It is not clear that the data was sampled in a way to
reveal a rapid pulsar signal, so this issue might still
be open.

The role of Poynting flux needs greater consideration,
both in the context of the formation of magnetars
and polarized supernovae and \grbs.  For perspective,
Usov (1999) has noted that a strong Poynting flux
would not allow differential motion of particles
and hence would tend to suppress internal shocks.

The interim conclusion is very clear.  We need to
explore the physics of rotating, magnetic core
collapse in considerably more detail than has
been done to date.

\begin{acknowledgments}  

This work is supported in part by NSF Grant 9818960, 
NASA grants LSTA-98-02 and HF-01085.01-96A and by a grant from the 
Texas Advanced Research Program.

\end{acknowledgments}


\begin{references}   


\bibitem[]{}
1. Blackman, E. G. \& Yi, I. 1998, ApJ, 498, L31 

\bibitem[]{}
2. Colgate, S. A. 1974, ApJ, 187, 333

\bibitem[]{}
3. Duncan, R. C. \& Thompson, C. 1992, ApJ, 392, L9       

\bibitem[]{}
4. M\"uller, E. \& Hillebrandt, W. 1979, A\&A, 80, 147

\bibitem[]{}
5. H\"oflich, P., Wheeler, J. C., \& Wang, L. 1999, ApJ, 521, 179  

\bibitem[]{}
6. H\"oflich, P., et al. 2000, in preparation 

\bibitem[]{}
7. Khokhlov, A.M., 1998, J. Comput. Phys., 143, 519    

\bibitem[]{}
8. Khokhlov  A.M., et al. 1999, ApJ, 524, L107

\bibitem[]{}
9. Kouveliotou, C., Strohmayer, T., Hurley, K., Van Paradijs, J., 
Finger, M. H., Dieters, S., Woods, P., Thompson, C. \& Duncan, R. C.
1998, ApJ, 510, 115 

\bibitem[]{}
10. LeBlanc, J. M. \& Wilson, J. R. 1970, ApJ, 161, 541

\bibitem[]{}
11. Symbalisty, E. M. D. 1984, ApJ, 285, 729
                                                  
\bibitem[]{}
12. Usov, V. V. 1999, astro-ph/9909435
 
\bibitem[]{}
13. Wang, L., Howell, D. A.  H\"oflich, P., \& Wheeler, J. C. 1999, 
ApJ submitted

\bibitem[]{}
14. Wang, L., Wheeler, J. C., Li, Z. W., \& Clocchiatti, A. 1996, ApJ, 467, 435
      
\bibitem[]{}
15. Wheeler, J. C.,  Yi, I., H\"oflich, P., \& Wang, L. 1999, ApJ, in press

\end{references}
\end{document}